%% file: main.tex
\documentclass[acmsmall,screen]{acmart}
\AtBeginDocument{%
  }

\setcopyright{none}
\author{Shiwen Shan}
\orcid{0009-0000-1317-8957}
\affiliation{%
  \institution{School of Software Engineering, Sun Yat-sen University} %School of Software Engineering, 
  \city{Zhuhai City}
  \country{China}
}
\email{shanshw@mail2.sysu.edu.cn}

\author{Yintong Huo}
\orcid{0009-0006-8798-5667}
\affiliation{%
  \institution{Singapore Management University}
  \country{Singapore}
}
\email{ythuo@smu.edu.sg}

\author{Zhiying Wu}
\orcid{0000-0003-4633-0008}
\affiliation{%
  \institution{School of Software Engineering, Sun Yat-sen University} %School of Software Engineering, 
  \city{Zhuhai City}
  \country{China}
}
\email{wuzhy286@mail.sysu.edu.cn}

\author{Yuxin Su}
\orcid{0000-0002-3338-8561}
\affiliation{%
    \institution{School of Software Engineering, Sun Yat-sen University}
  \city{Zhuhai City}
  \country{China}
}
\email{suyx35@mail.sysu.edu.cn}

\usepackage{collab}
\usepackage{tcolorbox}
\usepackage{booktabs}
\usepackage{array}
\usepackage{xspace}
\usepackage{algorithm}
\usepackage{algpseudocode}
\usepackage{graphicx}
\usepackage{listings}
\usepackage{colortbl}
\usepackage{multirow}
\collabAuthor{sw}{teal}{sw}

\def\tool{\texttt{StateMirror}\xspace}
\newcolumntype{L}[1]{>{\raggedright\arraybackslash}p{#1}}
\definecolor{main}{HTML}{b3b3b3}    % setting main color to be used
\definecolor{sub}{HTML}{f1f1f1}     % setting sub color to be used
\newtcolorbox{findingbox}[1]{
  title=\textbf{#1},
  colback=blue!3,
  colframe=blue!45!black,
  boxrule=0.45pt,
  arc=1.5pt,
  left=6pt,
  right=6pt,
  top=5pt,
  bottom=5pt
}

\begin{document}

\title{StateMirror: Pre-Execution Detection of Hallucinated Tool Calls in Coding Agents}



\begin{abstract}
Large language model (LLM)-based coding agents may propose invalid actions that are not applicable in the execution environment, such as accessing non-existent paths or missing packages.
We term this mismatch \textit{hallucinated tool use}.
A preliminary study of 4,398 agent trajectories reveals that these hallucinations are pervasive and costly, requiring a median of 3 additional interaction steps and 22.4K tokens to recover. Existing frameworks rely on reactive, post-execution feedback, wasting runtime resources.
Nevertheless, since an agent's execution environment is grounded in concrete codebase and harness components, many action prerequisites can be extracted and validated \textit{prior to} execution.

% per episode.
% While these verifiable filesystem and namespace mismatches present a clear opportunity for targeted detection, existing debugging frameworks rely on post-execution feedback, wasting substantial runtime resources.

To this end, we introduce \tool, a lightweight pre-execution detector.
\tool intercepts actions at the harness boundary, extracts state predicates via AST and command analysis, and deterministically validates them against the live repository.
By pinpointing exact environment mismatches before execution, \tool preempts invalid actions and subsequent recovery cycles.
Evaluation on 2,508 actions shows \tool achieves 0.940 precision and 0.817 F1 in detecting such hallucinations, surpassing the strongest LLM judge by 35.3\% in F1 while reducing median detection latency from 17.73 s to 1.50 ms with zero token consumption.
For practical mitigation, \tool's pre-execution feedback enables agents to rectify over 95\% of hallucinated tool calls.
During live mitigation on 79 tasks, \tool reduces agent execution time by 15.8\% and token consumption by 11.8\% while maintaining task resolution.
\end{abstract}
\maketitle

\input{Sections/01-intro}
\input{Sections/02-back}

\input{Sections/03-preliminary}
\input{Sections/04-StateMirror}

\input{Sections/05-Evaluation}

\input{Sections/06-case-study}
\input{Sections/related-work}
\input{Sections/07-threats}
\input{Sections/08-Conclusion}

% \section*{Data Availability}
% The artifact of \tool is available at 
% \url{https://anonymous.4open.science/r/StateMirror/}.

% \newpage
\bibliographystyle{ACM-Reference-Format}
\bibliography{main}

\end{document}

%% file: Sections/01-intro.tex
\section{Introduction}
Modern coding agents solve software engineering tasks by iteratively interacting with their environments~\cite{wang2024openhands,bouzenia2024repairagent}.
These agents typically follow the reasoning-and-acting (ReAct) paradigm~\cite{yao2022react}, using tool calls to inspect repository contents, modify files, and run tests. 
However, these actions often lack grounding in the workspace, leading to failures such as referencing invalid filesystem paths, missing modules, or undefined symbols. 
% Such errors, particularly hallucinated imports, are common in generated code and typically become visible only during execution~\cite{zhang2025codehallucination,wang2025codegenerrors}.

\begin{figure*}[t]
    \centering
    \includegraphics[width=\textwidth]{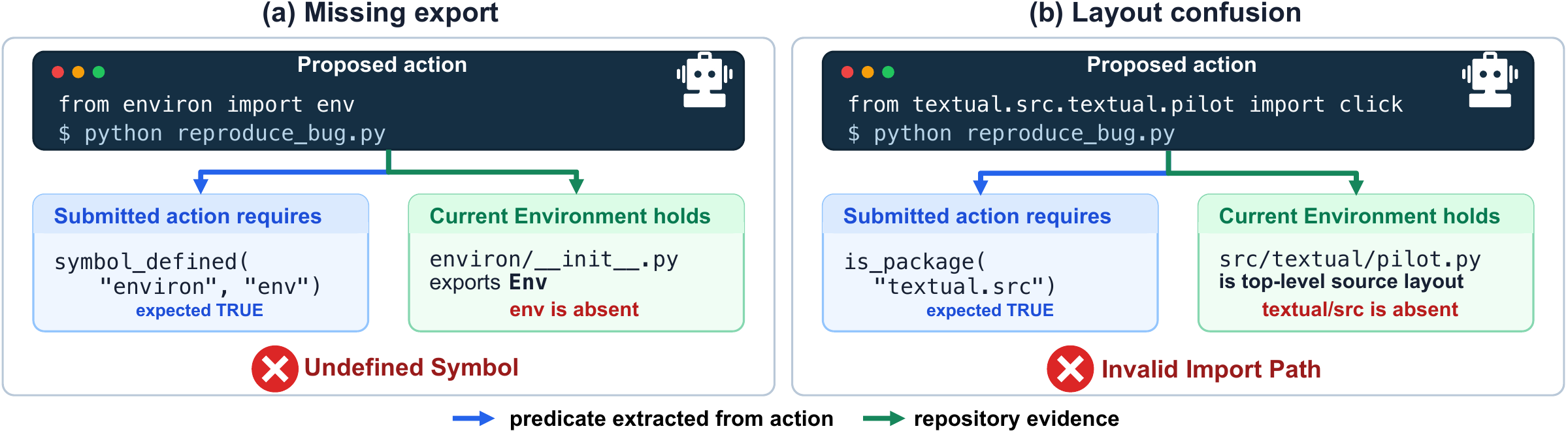}
    \caption{Examples of hallucinated tool calls. (a) The action imports \texttt{env}, but the module exports only \texttt{Env}. (b) The action imports \texttt{textual.src}, but the layout is \texttt{src/textual}.}
    \label{fig:examples}
\end{figure*}

We define this failure mode as \textbf{hallucinated tool use}, occurring when an agent's proposed action violates preconditions imposed by the actual environment~\cite{hoare1969axiomatic,meyer1992dbc}.
Unlike reasoning flaws or semantic code bugs that execute successfully but fail the task, hallucinated tool use embeds invalid references (e.g., non-existent entities) into tool calls, making them fundamentally unexecutable as intended. 
Figure~\ref{fig:examples} illustrates two hallucinated tool uses: (a) assumes an unexported symbol within a target namespace, (b) misinterprets the filesystem layout.
% Both actions violate deterministic workspace preconditions prior to execution.

To understand the impact of hallucinated tool use, we conduct a preliminary study analyzing 4,398 SWE-agent trajectories on SWE-bench-Extra~\cite{nebius2024swebenchextra,yang2024sweagent}, identifying 9,595 hallucinated tool calls across 120,801 total calls. 
Our study reveals three key findings.
First, hallucinated tool use is \textit{pervasive}, occurring in 46.6\% of failed and 32.3\% of resolved trajectories. 
Second, recovering from these hallucinations is \textit{costly}. When a hallucinated tool call fails (e.g., with a \texttt{ModuleNotFoundError}), the agent may take several steps before reaching a grounded retry, incurring a median recovery cost of 3 steps and 22.4K tokens.
Furthermore, 30.3\% of these hallucinations never reach a successful grounded retry.
Third, their root causes are \textit{concentrated}: filesystem mismatches (such as missing files or paths) account for 43.3\% of the cases, and namespace mismatches (such as missing modules or undefined symbols) account for 41.9\%. This concentration of verifiable error types indicates that most hallucinations can be validated before the step proceeds, thereby avoiding costly post-failure recovery.

However, existing approaches are not designed for such pre-execution grounding. 
They primarily target high-level inference flaws and often overlook tool hallucinations caused by inconsistencies with the execution environment. 
Some methods detect reasoning failures only through post-execution feedback~\cite{shinn2023reflexion,chen2023selfdebug,fan2023llmrepair}, incurring runtime overhead after the erroneous action has already been executed. Others rely on LLM-based judges over execution histories~\cite{zhang2026agentforesight}, which introduce additional inference cost and may suffer from reasoning biases~\cite{zheng2023llmjudge}.
% As our study identifies, filesystem and Python namespace assumptions account for 85.2\% of confirmed hallucinated tool calls.
% This concentration motivates lightweight, deterministic probes that check action requirements against the environment before execution, avoiding additional LLM inference.
% Despite this opportunity, existing approaches to improving agent reliability primarily rely on post-execution feedback loops, which execute an action, observe the result, and revise~\cite{shinn2023reflexion,chen2023selfdebug,fan2023llmrepair,jiang2023impact}.
% While these methods can resolve semantic errors that appear during execution, relying on such post-hoc feedback to address hallucinated tool use wastes substantial runtime resources.

To address this gap, we present \tool, a lightweight detector that catches proposed hallucinated tool use before it proceeds.
Because tool preconditions are grounded in verifiable environment artifacts, such as the repository codebase and runtime harness, we can validate them statically without executing the action.
Specifically, \tool intercepts actions at the harness boundary and verifies their preconditions through a two-step process. First, a predicate extractor statically parses the proposed action to derive typed state predicates, such as \texttt{importable} or \texttt{symbol\_defined}. Second, a read-only prober deterministically checks these predicates against the live execution environment. If any predicate is violated, \tool returns a typed warning detailing the exact mismatch. Operating deterministically without LLM invocation, \tool consumes no tokens, incurs minimal latency, and preempts costly post-execution recovery cycles.

We evaluate \tool on a benchmark of 2,508 expert-labeled actions from 1,197 public failed trajectories of four agent frameworks with different harness designs: SWE-agent, OpenHands, Sonar, and Honeycomb.
Following the annotation workflow of our preliminary study, we retrieve candidate actions via failure keywords and tracebacks, replay prior edits to reconstruct the exact pre-action workspace, and have annotators label each action independently.
\tool achieves a hallucination detection precision of 0.940 and an F1 score of 0.817, exceeding the strongest LLM judge by 0.213 F1 and a scaffold linter by 0.772 F1.
Furthermore, \tool is also fast and cheap to run, completing each detection in a median of 1.50\,ms with no token cost.

To show the practicality, we evaluate \tool on both immediate action repair and end-to-end task intervention.
For action repair, \tool's diagnostic result enables agents to rectify 95.0\% of hallucinated tool calls under controlled revision and 97.8\% during live SWE-agent execution.
In end-to-end intervention across 79 hallucination-affected tasks, \tool reduces execution time by 15.8\% and token consumption by 11.8\%, while maintaining task resolution.

\noindent\textbf{Contributions.} This paper makes three contributions:
% \begin{itemize}

$\bullet$ \textit{Problem Formulation.} We formalize the concept of hallucinated tool use, and empirically analyze 4,398 agent trajectories, characterizing its prevalence, costly recovery loops, and error patterns.

$\bullet$ \textit{Approach.} We present \tool, a deterministic pre-execution detector that derives state predicates and validates them
against the execution environment.

$\bullet$ \textit{Evaluation.} We show that \tool achieves 0.940 precision in hallucination detection at millisecond-level latency. Furthermore, its pre-execution feedback rectifies 95\% of invalid tool calls, reducing end-to-end task time by 15.8\% and token costs by 11.8\%.
% \end{itemize}

%% file: Sections/02-back.tex
\section{Background}

\subsection{ReAct-style Coding Agents}
Interactive coding agents complete software engineering tasks by combining reasoning and tool use.
This process typically follows the ReAct paradigm~\cite{yao2022react}.
In this loop, the agent generates a structured action, the environment executes it, and the harness returns an observation.
% Benchmarks like SWE-bench evaluate this interactive process using real GitHub issues~\cite{jimenez2023swebench}.
Various agent runtimes and repair systems operate within this paradigm by translating model outputs into environment commands~\cite{yang2024sweagent,bouzenia2024repairagent,zhang2024autocoderover}.
Additionally, prior studies investigate different formats for these actions, such as direct code execution or standardized JSON schemas~\cite{wang2024codeact,yang2023intercode}.
While earlier models focus on simple API invocation~\cite{schick2023toolformer,patil2024gorilla,qin2024toolllm}, software-engineering agents automate complex tasks like test generation and behavior recovery~\cite{ruan2025specrover,kang2023libro}.
However, integrating and running these agents in real-world systems remains highly challenging~\cite{chen2025llmchallenges, lu2025exploring}.
Our work focuses specifically on the tool-use loop of these coding agents.
We analyze the proposed tool calls and their arguments rather than the internal reasoning of the model.

% Interactive coding agents complete software engineering tasks by combining reasoning and tool use.
% This process typically follows the ReAct paradigm~\cite{yao2022react}.
% In this loop, the agent generates a structured action, the environment executes it, and the harness returns an observation.
% % Benchmarks like SWE-bench evaluate this interactive process using real GitHub issues~\cite{jimenez2023swebench}.
% Various agent runtimes and repair systems operate within this paradigm by translating model outputs into environment commands~\cite{yang2024sweagent,bouzenia2024repairagent,zhang2024autocoderover}.
% Additionally, prior studies investigate different formats for these actions, such as direct code execution or standardized JSON schemas~\cite{wang2024codeact,yang2023intercode}.
% While earlier models focus on simple API invocation~\cite{schick2023toolformer,patil2024gorilla,qin2024toolllm}, software-engineering agents automate complex tasks like test generation and behavior recovery~\cite{ruan2025specrover,kang2023libro}.
% However, integrating and running these agents in real-world systems remains highly challenging~\cite{chen2025llmchallenges}.
% Our work focuses specifically on the tool-use loop of these coding agents.
% We analyze the submitted tool calls and their arguments rather than the internal reasoning of the model.

\subsection{Preconditions and Runtime Checking}

In axiomatic semantics of programs, pre- and post-condition reasoning is concerned with checking that on any program input satisfying the precondition $P$, the program output should satisfy the postcondition $Q$. For example, the precondition $P$ may be that an input file exists or that some variable value is positive, and the postcondition $Q$ may specify the verification condition that we want to be satisfied. 
Pre- and post-condition checking underlies many program verification techniques, including static analysis tools such as Dafny~\cite{leino2010dafny} or Frama~\cite{cuoq2012frama} or dynamic analysis techniques such as runtime monitoring~\cite{leucker2009runtime}.

% Furthermore, rather than analyzing an abstraction of all possible states like abstract interpretation~\cite{cousot1977abstract}, \tool probes the concrete working environment. 
% This ensures that any violated predicate is reported as a definite contradiction rather than a potential risk.

%% file: Sections/03-preliminary.tex
\section{Preliminary Study}
We conduct a preliminary study of hallucinated tool calls in coding-agent trajectories, manually characterizing their prevalence, recovery costs, and root causes.

\subsection{Problem Statement}
% In our study, we formalize the execution environment and the interaction of actions with the repository workspace. 
To formalize the interaction between agent actions and the execution environment,
we define a state $s$ as the state of the workspace and harness context immediately prior to an action's execution. This state encompasses concrete, verifiable environment properties, including the file tree, active working directory, available packages, and exported symbols. 
Let $a$ denote a proposed structured action, and let $\phi(a)$ represent the set of predicates that action $a$ implicitly assumes to hold. We define the hallucinated tool use set $H(a, s)$ (or simply the hallucination set) as the set of such predicates that are false in the current environment:
\[
H(a, s) = \{ p \mid p \in \phi(a), \psi(p, s) = \mathrm{false} \}.
\]
Here, $\psi(p, s)$ is a probe function that evaluates the truth value of predicate $p$ within state $s$.
We classify a proposed action $a$ as a hallucinated tool call in state $s$ if $ H (a,s)\neq\emptyset$.
% Hallucinated tool use thus represents a deterministic contradiction in which the environment state falsifies the action's requirements before execution.
% It is distinct from a general logical defect in a patch because the action itself is structurally invalid and cannot execute as intended in the current workspace.

\subsection{Study Setup}
We conduct a pilot study to investigate the hallucinated tool use problem, followed by three research questions~(RQs):
(1) RQ1: Does hallucinated tool use exist in real coding-agent trajectories, and what is its prevalence?
(2) RQ2: What is the recovery cost of hallucinated tool calls in terms of agent steps and tokens?
(3) RQ3: What is the distribution of different types of hallucinated tool use?

\begin{figure}
    \centering
    \includegraphics[width=\linewidth]{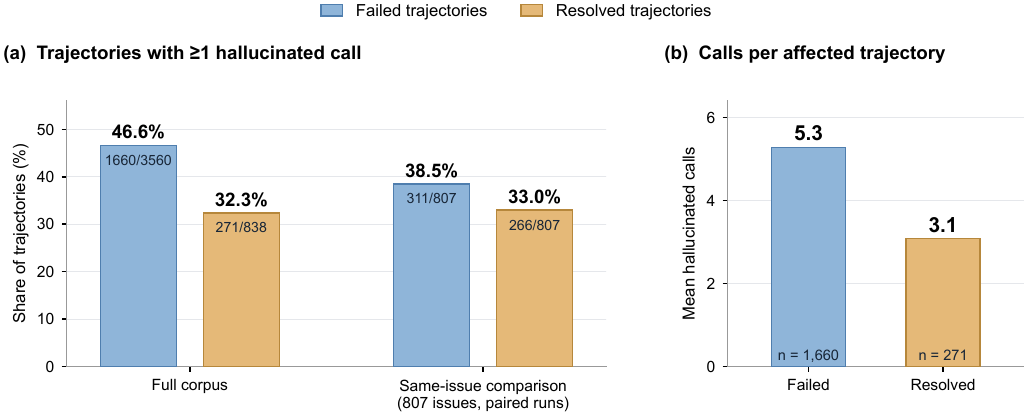}
    \caption{RQ1: Prevalence of hallucinated tool calls in failed and resolved agent trajectories.}
    \label{fig:prelim-summary}
\end{figure}
\subsubsection{Dataset and Ground Truth}
We analyze hallucinated tool use through a systematic three-stage workflow consisting of automated screening, human adjudication, and cost replay.

% First, during automated screening, we process a public corpus of 80,036 SWE-agent runs on SWE-bench~\cite{nebius2024sweagenttrajectories}, retaining one trajectory per issue and outcome to obtain 3,560 failed and 838 resolved runs.
We identify and validate hallucinated tool calls in two stages.
First, during automated screening, we analyze a public corpus of 80,036 SWE-agent runs~\cite{nebius2024sweagenttrajectories} on SWE-bench-Extra~\cite{nebius2024swebenchextra}, which spans a broader set of Python repositories than the original SWE-bench~\cite{jimenez2023swebench}. 
To compare behaviors across different outcomes, we retain one trajectory per issue and outcome, yielding 3,560 failed trajectories (with 107,308 actions) from 1,274 repositories and 838 resolved trajectories (with 13,493 actions) from 491 repositories.
From these trajectories, we identify candidate mismatches through two channels: (1) execution failures filtered by deterministic rules and (2) editor command errors rejected directly by the harness.
Second, in the human adjudication stage, two co-authors independently annotate these candidates by inspecting the exact workspace state and full trajectory history of each action, resolving disagreements through a third annotator.

In total, this process confirms 9,595 hallucinated tool calls in total (4,715 from execution failures and 4,880 from harness rejections), establishing a ground-truth dataset of hallucinated tool calls.

\subsection{Empirical Findings}
We present the empirical results of our study for each research question.

\subsubsection{RQ1: Existence and Prevalence}
We first measure the prevalence of hallucinated tool uses in real trajectories, defined as the proportion of trajectories that contain at least one hallucinated tool call, separately for failed and resolved runs.
% Measuring this frequency helps establish how often pre-execution checks are potentially applicable.

As shown in Figure~\ref{fig:prelim-summary}(a), 46.6\% of failed trajectories contain at least one hallucinated tool call, compared with 32.3\% of resolved trajectories; overall, 1,931 of the 4,398 trajectories (43.9\%) are affected. This trend also holds for issues that contain both failed and resolved runs. Among trajectories with hallucinated tool use, failed runs contain 5.3 such calls on average, compared with 3.1 in resolved runs (Figure~\ref{fig:prelim-summary}(b)).

The high prevalence across both failed and resolved runs shows that hallucinated tool use is not limited to unsuccessful trajectories and can arise regardless of the final task outcome. Moreover, hallucinated tool calls can recur within a trajectory, indicating that insufficient environment grounding may persist across agent execution. This creates a clear need to verify tool-use assumptions against the concrete environment before execution.

% As shown in Figure~\ref{fig:prelim-summary}(a) (left),
% our expert adjudication confirms 8,760 hallucinated tool calls across 1,660 of the 3,560 failed trajectories (46.6\%), spanning 766 repositories.
% In the 838 resolved trajectories, it confirms 835 hallucinated tool calls across 271 trajectories (32.3\%), spanning 207 repositories.
% Over all 4,398 trajectories, 1,931 (43.9\%) contain at least one hallucinated tool call; the first channel alone accounts for 1,178 (26.8\%), and 753 trajectories contain only harness-channel hallucinations.

% Because failed and resolved runs of different issues may differ in difficulty, we also compare the 807 issues for which the corpus contains both a failed and a resolved trajectory. 
% As Figure~\ref{fig:prelim-summary}(a) (right) shows, 311 of the failed and 266 of the resolved trajectories contain a hallucinated tool call.
% Hallucinated tool calls are thus somewhat less frequent in resolved runs, but far from absent.
% As Figure~\ref{fig:prelim-summary}(b) illustrates, failed trajectories also accumulate more hallucinations: among trajectories with at least one hallucinated tool call, failed runs contain 5.3 calls on average, compared with 3.1 in resolved runs.

\begin{tcolorbox}[
    colback = sub, 
    colframe = main, 
    boxrule = 0pt, 
    toprule = 1pt, % top rule weight
    bottomrule = 1pt, % bottom rule weight
    sharp corners
]
\vspace{-0.5em}
\small
\label{box:fd1}
\textbf{Finding \#1: Existence and Prevalence} \\
Hallucinated tool use is prevalent, affecting 46.6\% of failed trajectories and 32.3\% of resolved trajectories. Affected trajectories can also contain repeated hallucinated calls, indicating that insufficient environment grounding can persist during agent execution.
% and in similar numbers on issues with both outcomes (311 vs.\ 266 of 807). 
% Each hallucinated tool call is an action whose false precondition was checkable before execution.
\vspace{-0.5em}
\end{tcolorbox}

\subsubsection{RQ2: Recovery Cost of Hallucinated Tool Use}
To quantify the execution cost of hallucinated tool calls, we trace the 1,931 trajectories containing at least one confirmed hallucination. We define a \emph{hallucination episode} as the sequence of actions starting from an initial hallucinated tool call and spanning all subsequent attempts until \emph{recovery}. Recovery occurs when the agent successfully executes the intended command without error, correctly referencing the target entity (e.g., a module, symbol, or path extracted from the execution feedback).
We measure the episode cost in agent steps and tokens, using the cl100k tokenizer~\cite{openai2023tiktoken}. In total, we analyze 1,606 episodes originating from commands with an identifiable target entity. We exclude harness-channel hallucinations because agents rarely re-run the same editor command after a harness error.

Hallucinated tool calls impose substantial recovery costs. A recovered episode consumes a median of 3 agent steps and 22.4K tokens before a grounded retry. Across the 1,120 recovered episodes, these attempts consume 39.3M tokens in total. In addition, 30.3\% of episodes are never retried after the error; in these cases, agents proceed for a median of 18 subsequent steps and eventually fail the task. Thus, hallucinated tool calls add execution overhead, consuming agent steps and tokens and slowing task completion.

\begin{tcolorbox}[
    colback = sub, 
    colframe = main, 
    boxrule = 0pt, 
    toprule = 1pt, % top rule weight
    bottomrule = 1pt, % bottom rule weight
    sharp corners
]
\vspace{-0.5em}
\small
\label{box:fd2}
\textbf{Finding \#2: Recovery Cost of Hallucinated Tool Use} \\
Recovered hallucination episodes consume a median of 3 agent steps and 22.4K tokens before a grounded retry. Moreover, 30.3\% of episodes are never retried, showing that hallucinated tool use can impose persistent execution costs.
\vspace{-0.5em}
\end{tcolorbox}

\input{Tables/tab-prelim-family}
\subsubsection{RQ3: Type Distribution}
We next examine the types of assumptions violated by the 9,595 confirmed hallucinated tool calls. 
To derive the categories, two of the co-authors follow an open card-sorting procedure~\cite{spencer2009card,zimmermann2016card}: during adjudication, whenever an annotator marks a call as a hallucinated tool call, they also label the category of assumption it violates, introducing a new category when no existing one fits. The two annotators then iteratively compare, merge, and refine their categories into a shared set, as summarized in Table~\ref{tab:prelim-family}.
Over all 9,595 calls, the two annotators reach a Cohen's $\kappa$ of 0.869 (95.6\% raw agreement), and the third co-author resolves the remaining disagreements.

The distribution is highly concentrated in a small number of categories.
As shown in Table~\ref{tab:prelim-family}, filesystem and Python namespace assumptions account for more than 85\% of all confirmed hallucinated tool calls. They also affect a large fraction of trajectories, appearing in 58.2\% and 52.7\% of them, respectively. Harness assumptions are much less common overall and are concentrated in a small number of trajectories. Other command and runtime assumptions occur only rarely.

This concentration has an important implication. Most hallucinated tool calls arise from a small set of concrete properties of the execution environment, such as whether a file exists or whether a Python symbol is available. These properties can be represented as explicit predicates and checked against the execution environment before execution. The limited diversity of violated assumptions therefore makes systematic pre-execution detection practical.
% We then analyze the category distribution of the 9,595 confirmed hallucinated tool calls.
% Table~\ref{tab:prelim-family} groups the categories of hallucinated tool calls by the underlying assumptions they violate. Filesystem assumptions represent 43.3\% of all confirmed actions and appear in 58.2\% of the trajectories. Python namespace assumptions account for 41.9\% of the actions and appear in 52.7\% of the trajectories. 
% Together, these two classes cover over 85\% of the observed issues. Harness assumptions account for 14.7\% of the actions and appear in 11.7\% of the trajectories, while the command and runtime class is nearly empty. Harness-channel hallucinations are highly concentrated, as ten trajectories contribute 86.3\% of the schema violations. The distribution also varies slightly by trajectory outcome. Namespace assumptions make up 41.3\% of the confirmed actions in failed trajectories and 48.7\% in resolved trajectories. 
% These results show that a small set of well-defined predicates covers the vast majority of hallucinated tool calls, which makes the problem highly amenable to pre-execution checks.

\begin{tcolorbox}[
    colback = sub, 
    colframe = main, 
    boxrule = 0pt, 
    toprule = 1pt, % top rule weight
    bottomrule = 1pt, % bottom rule weight
    sharp corners
]
\vspace{-0.5em}
\small
\label{box:fd3}
\textbf{Finding \#3: Type Distribution} \\
More than 85\% of confirmed hallucinated tool calls involve filesystem or Python namespace assumptions. Their concentration in a small set of concrete environment properties makes pre-execution detection practical.
% The identified hallucinated tool calls are concentrated in a small set of typed predicates: filesystem and namespace assumptions each account for over 40\% of them, and harness assumptions for most of the rest.
\vspace{-0.5em}
\end{tcolorbox}

%% file: Tables/tab-prelim-family.tex
\begin{table}[t]
\centering
\caption{RQ3: Distribution of hallucinated tool use categories.
\emph{Channel} is the screening channel that found the category: \emph{Exec.} for execution failures, \emph{Harness} for editor-command error messages.}
%Trajectory counts are unique within each row; a trajectory can contain
% gaps from multiple categories.
\label{tab:prelim-family}
\small
\setlength{\tabcolsep}{3pt}
\begin{tabular}{@{}L{0.36\linewidth}lrrrr@{}}
\toprule
\textbf{Class / category} & \textbf{Channel} & \textbf{Failed} & \textbf{Resolved} & \textbf{Share} & \textbf{Traj.} \\
\midrule
\textbf{Filesystem and cwd state} & & \textbf{3,762} & \textbf{395} & \textbf{43.33\%} & \textbf{1,123} \\
\quad File, path, or cwd target & Exec. & 613 & 75 & 7.17\% & 249 \\
\quad Editor path operand & Harness & 3,149 & 320 & 36.15\% & 946 \\
\addlinespace
\textbf{Python namespace and package layout} & & \textbf{3,616} & \textbf{407} & \textbf{41.93\%} & \textbf{1,017} \\
\quad Import or package availability & Exec. & 2,101 & 245 & 24.45\% & 689 \\
\quad Runtime symbol or API assumption & Exec. & 1,515 & 162 & 17.48\% & 471 \\
\addlinespace
\textbf{Harness and tool schemas} & & \textbf{1,378} & \textbf{33} & \textbf{14.71\%} & \textbf{225} \\
\quad Tool-call schema & Harness & 1,206 & 14 & 12.71\% & 101 \\
\quad Editor state (open file, line range) & Harness & 172 & 19 & 1.99\% & 131 \\
\addlinespace
\textbf{Command, test, and runtime state} & & \textbf{4} & \textbf{0} & \textbf{0.04\%} & \textbf{1} \\
\quad Dependency or runtime version & Exec. & 4 & 0 & 0.04\% & 1 \\
\midrule
\textbf{Total} & & \textbf{8,760} & \textbf{835} & \textbf{100.00\%} & \textbf{1,931} \\
\bottomrule
\end{tabular}
\end{table}

%% file: Sections/04-StateMirror.tex
\section{Methodology}
\subsection{Overview}

% \tool is a lightweight detector designed to assess action feasibility \textit{before} execution. As shown in Figure~\ref{fig:overview}, \tool intercepts each structured action proposed to the harness. It extracts the implicit assumptions, represented as predicates, and verifies them against the current workspace and harness state. A violated predicate indicates hallucinated tool use.

\begin{figure}[t]
    \centering
    \includegraphics[width=\linewidth]{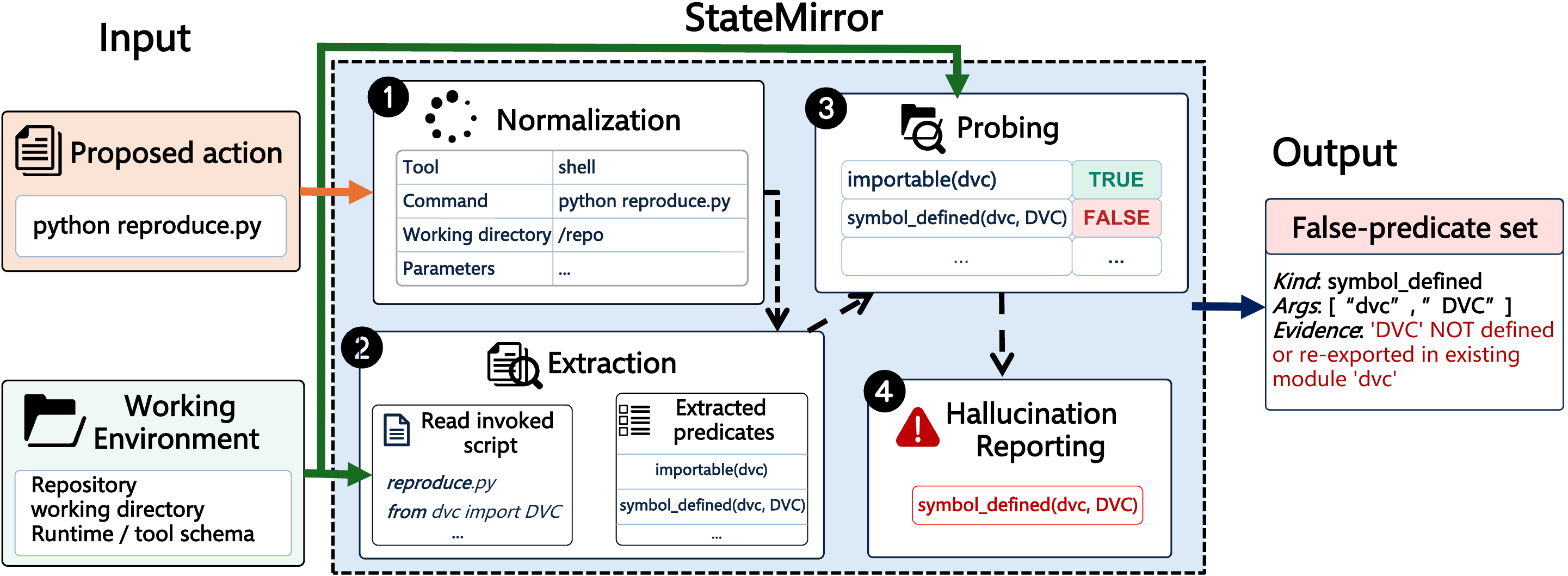}
    \caption{Overview of \tool. Predicates extracted from the proposed action are checked against the live environment before execution. Only those whose probes return \textsc{false} are reported as hallucinated tool use.}
    \Description{A flowchart with two inputs, a proposed action and the environment state, feeding StateMirror. The auditor extracts action-side predicates and probes the environment, then compares the two. A mismatch produces a false-predicate set containing only predicates whose probe is false.}
    \label{fig:overview}
\end{figure}

\tool is a lightweight detector designed to assess action feasibility \textit{before} execution. 
As shown in Figure~\ref{fig:overview}, \tool includes four stages: normalization, predicate extraction, environment probing, and reporting.
The extractor depends on action syntax, and the prober depends on the workspace and runtime artifacts available at check time.
Both components can be extended for new tools or environments without changing these stages.
% Figure~\ref{fig:overview} shows how \tool checks each proposed action before the harness executes it.
% Each structured action carries implicit assumptions about the working environment.
% \tool makes these assumptions explicit as predicates and checks them against the environment using probes.
% When a probe shows that an assumption does not hold, \tool records that predicate in a \emph{false-predicate set} and reports a warning.
% A non-empty set identifies a hallucinated tool call.

% Motivated by Finding~\#3, we organize these predicates into four families: filesystem, Python namespace, harness, and command/runtime predicates, as summarized in Figure~\ref{fig:method-families}. This design follows the empirical observation that hallucinated tool calls are concentrated in a small set of environment assumptions, with filesystem and Python namespace assumptions alone covering more than 85\% of the observed cases. The remaining two families capture harness-level constraints and command or runtime requirements. Together, these four families cover the main types of environment properties observed in our study.

% % \subsection{Action Normalization}
% % To enable effective and consistent detection across diverse agent actions, \tool{} first normalizes the proposed action into a structured representation. This provides a unified understanding of the action's intended execution, regardless of its original format (e.g., shell command, Python code). 
% % It captures essential information such as the action's kind, current working directory (cwd), structured parameters, command text, and directly embedded executable code, thereby exposing the concrete execution intent without inferring private model reasoning.

% \subsection{Detection Pipeline}

\subsection{Normalization}
To handle different forms of agent actions, such as shell commands and structured tool calls, \tool normalizes each proposed action into a unified representation for subsequent analysis.
This representation captures the action type, current working directory, tool parameters, shell command, and associated Python code.
For shell commands that invoke Python scripts (e.g., \texttt{python reproduce.py}), \tool reads the script from the current workspace and includes its source code in the normalized representation when available.
This allows subsequent predicate extraction to analyze the code that the harness would execute.

\subsection{Extraction \& Probing}
Hallucinated tool use occurs when a proposed action assumes environmental conditions that do not hold in the actual execution environment.
To detect such mismatches, \tool extracts the implicit assumptions from proposed actions as explicit predicates and verifies them against the current environment by probing into the environment.

\subsubsection{Extraction.}
Given a normalized action, \tool extracts typed predicates that represent the environmental conditions required for its execution.
Depending on the action type, the extractor analyzes shell commands, structured tool parameters, and Python code.
For Python code, it parses the source into an abstract syntax tree (AST) on demand to identify relevant operations and their dependencies.
Each extracted predicate specifies a concrete requirement, such as the availability of a Python module, the existence of a filesystem path, or the validity of a tool argument.
When a required value cannot be determined statically, the corresponding predicate is omitted.

\subsubsection{Probing.}
For each extracted predicate, \tool performs read-only checks against the current execution environment.
Depending on the predicate type, these checks inspect the repository structure, Python module definitions, filesystem metadata, runtime information, or harness schemas.
A predicate is considered violated when the corresponding environmental requirement is confirmed to be false.
When a check cannot reliably determine whether a requirement holds, \tool conservatively allows the predicate to pass and records the uncertainty in the probe evidence.

\subsubsection{Predicate Families.}
\begin{figure}[t]
    \centering
    \includegraphics[width=\textwidth]{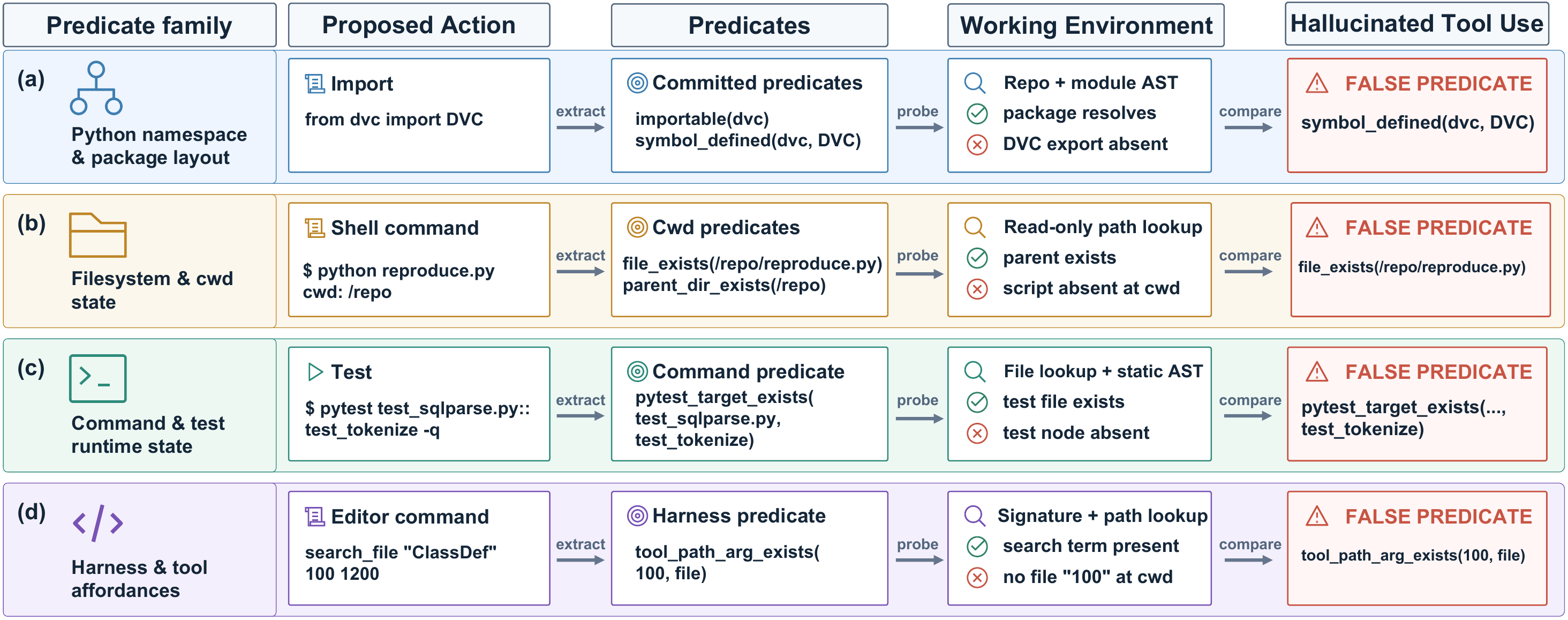}
    \caption{Four predicate families. Each panel
    starts from the proposed action, extracts the required predicates, checks them against the live environment using deterministic probes, and reports any contradiction before execution. 
    The examples illustrate assumptions regarding (a) namespace and package layout,
    (b) filesystem and working directory, (c) command, test, and runtime environment, and
    (d) harness and tool schemas.}
    % \Description{A four-panel comparison of StateMirror's predicate families.
    % Each panel shows the proposed action, extracted predicates, deterministic
    % state probes, and resulting typed hallucination. The panels cover Python namespace,
    % filesystem and working directory, command and runtime, and harness schema
    % state.}
    \label{fig:method-families}
\end{figure}

Motivated by Finding~\#3, we organize the extracted predicates into four families: Python namespace and package layout, filesystem and working directory, command and runtime state, and harness and tool schemas.
As shown in Figure~\ref{fig:method-families}, these families capture the main types of environmental assumptions identified in our preliminary study.
Each family defines specific predicates and corresponding checks to verify these assumptions against the execution environment.
\tool also supports extensions to new tools and environments by introducing additional predicates and their corresponding extractors and probers.

$\bullet$ \textbf{Family I: Python Namespace and Package Layout.}
This family verifies assumptions about Python modules, packages, and exported symbols.
Verifying these assumptions is important because missing modules or symbols can cause reproduction scripts and test helpers to fail during execution~\cite{kang2023libro}.

Since a module may exist without exposing a requested symbol, \tool examines both module availability and namespace contents, including symbols exposed through package re-exports.
To identify these requirements, \tool parses the AST of the Python code included in the normalized action and extracts predicates such as \texttt{importable}, \texttt{symbol\_defined}, and \texttt{is\_package}.
It then verifies these predicates against the repository structure, standard library, and installed packages by inspecting module definitions and common re-export patterns (e.g., \texttt{\_\_init\_\_.py}).
To limit probing overhead, \tool restricts dependency analysis to imports directly referenced in the proposed action or its invoked scripts.
Predicates involving names that cannot be resolved statically are omitted.

Figure~\ref{fig:method-families}(a) illustrates an example involving a missing Python symbol.
The agent proposes a script containing \texttt{from dvc import DVC}.
From its AST, \tool extracts \texttt{importable(dvc)} and \texttt{symbol\_defined(dvc, DVC)}.
The prober confirms that the \texttt{dvc} package exists but finds no definition or re-export of \texttt{DVC} in \texttt{dvc/\_\_init\_\_.py}.
It therefore evaluates \texttt{symbol\_defined(dvc, DVC)} as \textsc{false}.
This violated predicate identifies a hallucinated tool call that would otherwise result in an \texttt{ImportError}.

$\bullet$ \textbf{Family II: Filesystem and Current-Working-Directory State.}
This family verifies assumptions about filesystem paths and the current working directory (cwd).
Such verification is important because agents may repeatedly attempt to access nonexistent paths, wasting execution steps and computational resources on unsuccessful operations and subsequent recovery.

To identify these mismatches, \tool considers both the current cwd and the type of file operation (e.g., reading, writing, or executing a file).
Specifically, relative paths are resolved against the current cwd, while different operations impose different requirements.
For example, reading or executing a file requires the target to exist, whereas creating a file generally requires an existing parent directory.
Accordingly, \tool extracts path arguments from shell commands, Python file operations, and editor actions, generating operation-specific predicates such as \texttt{file\_exists} and \texttt{parent\_dir\_exists}.
The prober then checks whether the target files or directories exist at the resolved paths.
When a path cannot be determined statically, the corresponding predicate is omitted.

Figure~\ref{fig:method-families}(b) illustrates a missing script in the current working directory.
The agent proposes \texttt{python reproduce.py} with \texttt{/repo} as its cwd.
\tool resolves the script path and extracts \texttt{file\_exists(/repo/reproduce.py)} and \texttt{parent\_dir\_exists(/repo)}.
The prober confirms that \texttt{/repo} exists but finds no \texttt{reproduce.py} at the expected location.
Consequently, the first extracted predicate evaluates to \textsc{false}, identifying a hallucinated tool call that would otherwise fail with a \texttt{FileNotFoundError}.

$\bullet$ \textbf{Family III: Command, Test, and Runtime State.}
This family verifies assumptions about command availability, test targets, and runtime requirements.
Checking these assumptions before execution can prevent unnecessary command attempts and test runs, reducing the time and resources spent on diagnosing and correcting invalid actions.

To detect these mismatches, \tool analyzes shell commands and command-related tool parameters to extract executable names, pytest targets, and explicit runtime-version requirements.
It represents these requirements using predicates such as \texttt{executable\_available}, \texttt{pytest\_target\_exists}, and \texttt{runtime\_version\_satisfies}.
The prober then checks executable availability and runtime-version information in the current environment.
For pytest targets, it examines the target file and its static AST nodes to check whether the requested test exists.

Figure~\ref{fig:method-families}(c) illustrates a missing pytest target case.
The agent proposes an action, \texttt{pytest test\_sqlparse.py::test\_tokenize -q}, to verify a fix.
From the command, \tool extracts \texttt{pytest\_target\_exists(test\_sqlparse.py, test\_tokenize)}.
The prober confirms that \texttt{test\_sqlparse.py} exists and inspects its AST for the requested test node.
Since no matching function or method is found, the predicate evaluates to \textsc{false}.
\tool thus identifies the invalid test target before the command is executed.

$\bullet$ \textbf{Family IV: Harness and Tool Schemas.}
This family verifies assumptions about tool argument schemas and the current harness state.
Invalid tool arguments or unmet harness requirements may cause tool calls to be rejected, resulting in additional agent steps and recovery costs.

To identify these mismatches, \tool examines tool argument constraints (e.g., argument counts, types, and formats) and conditions maintained by the harness (e.g., whether a file is currently open and whether a specified line number is valid).
Accordingly, \tool extracts schema predicates, such as \texttt{tool\_args\_valid}, and condition-dependent predicates, including \texttt{tool\_path\_arg\_exists}, \texttt{editor\_file\_open}, \texttt{editor\_line\_in\_range}, and \texttt{path\_absent}.
The prober then checks these predicates against tool signatures, the filesystem, and information provided by the harness.
For file editing tools, \tool identifies the currently open file from the preceding tool response.
When the required information is unavailable, the corresponding predicate is omitted.

Figure~\ref{fig:method-families}(d) illustrates a tool argument mismatch.
The agent issues \texttt{search\_file "ClassDef" 100 1200}, intending to search lines 100--1200.
However, the tool signature \texttt{search\_file <term> [<file>]} interprets \texttt{100} as the file argument.
Accordingly, \tool extracts the predicate \texttt{tool\_path\_arg\_exists(100, file)}.
The prober finds no file named \texttt{100} in the current working directory, causing the predicate to evaluate to \textsc{false}.
This result identifies a hallucinated tool call that the harness would otherwise reject due to an invalid file argument.

\subsection{Hallucination Reporting}
After probing, \tool collects predicates whose probes return \textsc into a false-predicate set.
If the set is non-empty, the proposed action is identified as a hallucinated tool call.
Each violated predicate is reported as a typed warning that describes the mismatch between the proposed action and the actual environment.

%% file: Sections/05-Evaluation.tex
\section{Evaluation}

We evaluate the performance of \tool by answering the following RQs:
% \begin{itemize}

$\bullet$ \textbf{RQ4:} How accurately does \tool detect hallucinated tool calls?

$\bullet$ \textbf{RQ5:} What latency and token cost does \tool introduce?

$\bullet$ \textbf{RQ6:} How do different predicate families contribute to the overall detection accuracy?

$\bullet$ \textbf{RQ7:} How does bounded multi-hop scope affect the effectiveness and latency?

$\bullet$ \textbf{RQ8:} How effectively does \tool{} mitigate hallucinated tool calls within agent workflows?% \end{itemize}

\subsection{Environment and Implementation}
We implement \tool as a Python toolkit (5.8K lines of code) that normalizes actions and checks predicates using deterministic, read-only probes. It operates without executing actions or consuming LLM tokens. For runtime evaluation, we integrate it with SWE-agent~\cite{yang2024sweagent}. All experiments run on an Apple M4 Max machine with 36~GB RAM using Docker, executing tasks within their official SWE-bench containers. For offline evaluation, we reconstruct the pre-action workspace by replaying preceding edits and withhold all execution outcomes. We access all LLMs through the OpenRouter API~\cite{openrouter2024}, set the temperature to 0, and use direct import checking by default. 

% We implement \tool as an extensible Python toolkit (5.2K LOC, plus 0.6K LOC for the SWE-agent adapter) using standard \texttt{ast}, \texttt{shlex}, and \texttt{pathlib} modules to normalize actions.
% An extensible registry pairs each predicate extractor~$\phi$ with a deterministic, read-only probe~$\psi$ over cached AST and filesystem state, operating without executing actions or consuming LLM tokens.
% For runtime evaluation, we integrate \tool with SWE-agent~\cite{yang2024sweagent} v1.1.0 via an \texttt{on\_actions\_generated} hook to support report-only monitoring and pre-execution intervention.

% Experiments run on a MacBook Pro (Apple M4 Max, 14 cores, 36~GB RAM) using Docker~29.4.3 and SWE-ReX~1.4.0, executing each task within its official SWE-bench container.
% For offline evaluation, we faithfully reconstruct the pre-action workspace by replaying preceding file edits from base commits, strictly withholding execution outcomes and ground-truth labels from all evaluated methods.
% We set the LLM temperature to 0 for reproducibility and adopt direct (one-hop) checking by default.
% All prompt templates, model configurations, and execution scripts are available in our replication package.

\input{Tables/tab-evaluation-dataset}
\subsection{Dataset}
\label{sec:evaluation-dataset}

The preliminary study establishes a human-adjudicated workflow for recognizing pre-execution hallucinated tool calls in SWE-agent failures.
We extend this workflow to publicly available failed trajectories from SWE-agent~\cite{yang2024sweagent}, OpenHands~\cite{wang2024openhands}, Sonar~\cite{shariffdeen2025sonar}, and Honeycomb~\cite{honeycomb2024}, constructing a benchmark with expert-labeled positives and negatives.
Unlike the preliminary study, we build the benchmark exclusively from failed trajectories. Because a hallucinated tool call is defined by the mismatch between the action and its pre-action workspace rather than the final task outcome, its label remains independent of task resolution. We focus on failed runs because execution-failure evidence naturally concentrates there, and our public data sources (OpenHands, Sonar, and Honeycomb) primarily provide failed trajectories. To ensure comprehensive evaluation, RQ4 also separately reports the performance on resolved-run instances collected during the preliminary study.
% Ground-truth annotation is an offline, post-hoc process where human reviewers examine the submitted action, its subsequent observation, and the pre-action repository state to determine whether a hallucinated tool call occurred.
% In contrast, all evaluated detection approaches, including \tool and comparison baselines, operate strictly at the pre-execution boundary, receiving only the submitted action, the reconstructed pre-action workspace, and harness schemas without access to execution outcomes.

Specifically, we harvest 2,687 candidate failure actions across the four agent sources through automated keyword and traceback retrieval.
Human annotators independently inspect each candidate to identify hallucinated tool calls, resolving discrepancies through consensus adjudication.
To ensure state integrity, we replay prior prefix edits from the base repository commit to reconstruct the workspace immediately before each action.
We filter out 12 cases with unrecoverable execution environments and 167 candidate actions whose stated predicates hold in the reconstructed state, ensuring each retained positive represents a genuine pre-action hallucinated tool call.

Table~\ref{tab:evaluation-dataset} summarizes the resulting benchmark, which comprises 2,508 actions across 1,197 trajectories.
As shown in the table, positive hallucinations span five families across the four evaluated agent sources.
To prevent evaluation bias, we retain all identified hallucinated tool calls regardless of whether they fall within \tool's current predicate vocabulary, treating unsupported categories strictly as false negatives.
% Source- and family-level outcomes are reported with RQ1, rather than as separate construction tables.

\subsection{Metrics}

We evaluate \tool using the following metrics:
% \begin{itemize}

$\bullet$ \textbf{Detection Effectiveness:} Precision, recall, F1-score, and Accuracy against ground-truth labels to evaluate hallucination detection performance.

$\bullet$ \textbf{Computational Efficiency:} Execution latency (mean, median, p95, p99, and max) and token consumption to quantify pre-execution detection overhead.

$\bullet$ \textbf{Mitigation Performance:} We measure two aspects: (1) action repair rate, which is the proportion of warned actions successfully replaced by the agent to clear the gap, 
and (2) intervention efficiency, which measures the savings in agent time, steps, and tokens per task achieved by blocking actions that would otherwise fail on execution.
We also report the task resolve rate to verify that these savings preserve task success.
% relative to report mode, with official SWE-bench outcomes checking that intervention does not reduce task success.

\subsection{Baselines}
To benchmark the detection performance and efficiency of \tool, we compare it against three baseline families representing alternative pre-execution detection paradigms:
% \begin{itemize}

$\bullet$ \textbf{LLM Judges:} State-of-the-art reasoning models (Qwen3.7 Plus~\cite{qwen2026qwen37plus}, Kimi K3~\cite{moonshot2026kimik3}, and Codex 5.6 Sol~\cite{openai2026gpt56}) configured to predict hallucinated tool calls via zero-shot prompting in the LLM-as-a-judge setting~\cite{zheng2023llmjudge}. They receive the pre-action repository context, prefix-edit summaries, and submitted actions, but no execution outcomes.

$\bullet$ \textbf{Action-only:} An ungrounded baseline that flags all extracted predicates from the action as hallucinations without checking the workspace, isolating the necessity of repository-level probing.

$\bullet$ \textbf{Scaffold Linter:} A static linting baseline that inspects generated code for local construction-level errors (e.g., undefined names) but does not resolve imports against the workspace.

% All baselines are evaluated on the 2,508 benchmark actions under identical pre-execution conditions.
% The action-only and scaffold-linter baselines are deterministic and parameter-free.
% Each LLM judge operates with fixed prompt templates and model hyperparameters without access to expert labels, runtime feedback, or \tool diagnostics; unavailable API responses are conservatively treated as negative predictions.

\input{Sections/RQs/RQ1}
\input{Sections/RQs/RQ2}
\input{Sections/RQs/RQ3}
\input{Sections/RQs/RQ4}
\input{Sections/RQs/RQ5}

%% file: Tables/tab-evaluation-dataset.tex
% chktex-file 8
% chktex-file 24
% chktex-file 36
\begin{table}[t]
\centering
\caption{Composition of the action-level evaluation benchmark.
The benchmark contains 2,508 actions across 1,197 trajectories
(1,401 positives and 1,107 negatives).}
\label{tab:evaluation-dataset}
\small
\setlength{\tabcolsep}{2pt}
\begin{minipage}[t]{0.56\linewidth}
\centering
\textbf{(a) Benchmark by agent source}\\[0.35em]
\begin{tabular}{@{}lrrrr@{}}
\toprule
\textbf{Source} & \textbf{Traj.} & \textbf{Actions} & \textbf{Pos.} & \textbf{Neg.} \\
\midrule
SWE-agent & 325 & 758 & 500 & 258 \\
Honeycomb & 416 & 853 & 570 & 283 \\
Sonar & 316 & 490 & 205 & 285 \\
OpenHands & 140 & 407 & 126 & 281 \\
\midrule
Total & 1,197 & 2,508 & 1,401 & 1,107 \\
\bottomrule
\end{tabular}
\end{minipage}
\hfill
\begin{minipage}[t]{0.40\linewidth}
\centering
\textbf{(b) Positives by hallucination family}\\[0.35em]
\begin{tabular}{@{}lr@{}}
\toprule
\textbf{Family} & \textbf{Positives} \\
\midrule
Import & 785 \\
File/path & 311 \\
Symbol/export & 243 \\
Tool/harness schema & 49 \\
Environment & 13 \\
\midrule
Total & 1,401 \\
\bottomrule
\end{tabular}
\end{minipage}
\end{table}

%% file: Sections/RQs/RQ1.tex
\subsection{RQ4: Effectiveness Analysis}
\label{sec:rq1}

\input{Tables/tab-rq1-main}

Table~\ref{tab:rq1-main} compares \tool against baseline approaches.
We observe that conventional approaches struggle with the trade-off between coverage and grounding.
The action-only baseline achieves decent recall but produces high false positives because it inspects proposed commands only; without checking the underlying repository, it flags valid actions simply because they look unusual.
The scaffold linter yields a modest 0.045 F1 score because it only checks command syntax failures. Hallucinated tool calls are often syntactically valid but reference nonexistent runtime entities.
% The action-only baseline attains high recall but incurs 995 false positives due to the absence of repository-level probing, while the scaffold linter yields a negligible 0.045 F1 by checking only local syntax.
Although LLM judges outperform static linting, they struggle to infer full repository state from condensed context snapshots, with the strongest model (Qwen3.7 Plus) reaching only 0.604 F1.
In contrast, \tool achieves 0.940 precision, 0.817 F1, and 0.819 accuracy, outperforming Qwen3.7 Plus by 35.3\% in F1 and the most accurate baseline (Kimi K3, 0.616) by 33.0\% in accuracy. These results demonstrate that effective hallucination detection requires deterministic probing against the live environment.

\input{Tables/tab-rq1-robustness-errors}

We further analyze the failure cases of \tool, as detailed in Table~\ref{tab:rq1-robustness-errors}. 
The 390 false negatives are heavily concentrated in the pre-execution action extraction layer rather than state probing logic, with code extraction, command extraction, and selected-file reconstruction accounting for 73.8\% (288/390) of misses.
These misses primarily stem from implicit preconditions omitted by the action extractors, including library-call arguments, CLI options, and unmaterialized test files, which account for Honeycomb's higher FN count (209 misses).
Across predicate families, symbol predicates achieve the highest recall (0.802), while import (0.703) and file/path (0.736) capture the vast majority of positive hallucinated tool calls.
Conversely, false positives remain low across all sources (64 total, spanning 6 to 31 per source), predominantly caused by files generated at run time by the action or an earlier step, path-like string literals, and transient environment state absent during offline static reconstruction.
% We retain a conservative, state-grounded decision rule, as naively treating unresolved external imports as valid would suppress genuine gap detections and compromise precision.

\subsubsection{Hallucinated tool calls from resolved runs.}
While our primary evaluation relies on the main benchmark of failed trajectories, evaluating \tool{} on successful runs is crucial to assess its generalizability. We therefore apply \tool{} to the resolved stratum from our preliminary study. This dataset contains 823 expert-confirmed hallucinated tool calls as positives and 43 rejected candidates as hard negatives.
On this stratum, \tool{} achieves 0.989 precision, 0.532 recall, and 0.692 F1, detecting at least one hallucinated tool call in 81.5\% of the trajectories. The lower recall compared to the main benchmark is primarily due to dynamic runtime attribute errors and temporary files created during execution, which fall outside our static predicate vocabulary. When excluding these dynamic cases, \tool{} detects the remaining static, in-vocabulary screened hallucinations with 79.1\% recall (254/321). Similarly, it achieves 92.1\% recall (164/178) on harness-error calls when excluding run-time file creation retries.

\begin{tcolorbox}[
    colback = sub,
    colframe = main,
    boxrule = 0pt,
    toprule = 1pt, % top rule weight
    bottomrule = 1pt, % bottom rule weight
    sharp corners
]
\vspace{-0.5em}
\small
\label{box:rq1}
\textbf{Answer to RQ4:} \tool achieves 0.940 precision, 0.722 recall, and an F1-score of 0.817, outperforming the strongest LLM judge by 35.3\% and the scaffold linter by 1,715.6\% in F1.
\vspace{-0.5em}
\end{tcolorbox}

%% file: Tables/tab-rq1-main.tex
\begin{table}[t]
\centering
\caption{RQ4: Static detection performance across 2,508 benchmark actions.
Bold and underline indicate the best and second-best results, respectively.}
\label{tab:rq1-main}
\scriptsize
\setlength{\tabcolsep}{1.8pt}
% Relative gain of StateMirror vs. the cell baseline: (ours - base) / base.
\providecommand{\rqonegoain}[1]{\textcolor{red}{\scriptsize+#1\%}}
\providecommand{\rqoneloss}[1]{\textcolor{blue!60!black}{\scriptsize#1\%}}
\resizebox{\columnwidth}{!}{%
\begin{tabular}{lrrrr|rr|rr|rr|rr}
\toprule
\textbf{Method} & \textbf{TP} & \textbf{FP} & \textbf{FN} & \textbf{TN} &
\textbf{P} & $\boldsymbol{\Delta}$ &
\textbf{R} & $\boldsymbol{\Delta}$ &
\textbf{F1} & $\boldsymbol{\Delta}$ &
\textbf{Acc} & $\boldsymbol{\Delta}$ \\
\midrule
Scaffold linter & 35 & \underline{131} & 1366 & 976 &
0.211 & \rqonegoain{345.5} &
0.025 & \rqonegoain{2788.0} &
0.045 & \rqonegoain{1715.6} &
0.403 & \rqonegoain{103.2} \\
\addlinespace[1pt]
\multicolumn{13}{l}{LLM judges} \\
\quad Qwen3.7 Plus & 749 & 330 & 652 & 777 &
0.694 & \rqonegoain{35.4} &
0.535 & \rqonegoain{35.0} &
0.604 & \rqonegoain{35.3} &
0.608 & \rqonegoain{34.7} \\
\quad Codex 5.6 Sol & 759 & 427 & 642 & 680 &
0.640 & \rqonegoain{46.9} &
0.542 & \rqonegoain{33.2} &
0.587 & \rqonegoain{39.2} &
0.574 & \rqonegoain{42.7} \\
\quad Kimi K3 & 624 & 185 & 777 & \underline{922} &
\underline{0.771} & \rqonegoain{21.9} &
0.445 & \rqonegoain{62.2} &
0.565 & \rqonegoain{44.6} &
\underline{0.616} & \rqonegoain{33.0} \\
\addlinespace[1pt]
Action-only & \textbf{1371} & 995 & \textbf{30} & 112 &
0.579 & \rqonegoain{62.3} &
\textbf{0.979} & \rqoneloss{-26.3} &
\underline{0.728} & \rqonegoain{12.2} &
0.591 & \rqonegoain{38.6} \\
\midrule
\tool~(Ours) & \underline{1011} & \textbf{64} & \underline{390} & \textbf{1043} &
\textbf{0.940} & --- &
\underline{0.722} & --- &
\textbf{0.817} & --- &
\textbf{0.819} & --- \\
\bottomrule
\end{tabular}%
}
\end{table}

%% file: Tables/tab-rq1-robustness-errors.tex
\begin{table*}[t]
\centering
\caption{RQ4: Failure analysis partitioning the 1,401 positive actions and 390 misses across agent sources, predicate families, and error layers.}
\label{tab:rq1-robustness-errors}
\footnotesize
% \begin{tabular*}{\textwidth}{@{\extracolsep{\fill}}lrlr@{}}
% % \toprule
% % Held-out predicate targets detected & 186/213 (0.873) & Synthetic import-control fires & 0/623 \\
% % Valid-source repo-internal fires & 100/124,814 (0.080\%) & Valid-source coverage & 8,070 files / 144 repos \\
% % \bottomrule
% \end{tabular*}
% \vspace{5pt}

\begin{minipage}[t]{0.3\textwidth}
\centering
\textbf{By agent source}\\[2pt]
\begin{tabular}{@{}lrrr@{}}
\toprule
Source & FP & FN & F1 \\
\midrule
Honeycomb & 6 & 209 & 0.771 \\
OpenHands & 18 & 38 & 0.759 \\
Sonar & 9 & 72 & 0.767 \\
SWE-agent & 31 & 71 & 0.894 \\
\midrule
Total & 64 & 390 & 0.817 \\
\bottomrule
\end{tabular}
\end{minipage}\hfill
\begin{minipage}[t]{0.4\textwidth}
\centering
\textbf{By positive family}\\[2pt]
\begin{tabular}{@{}lrrr@{}}
\toprule
Family & $N_+$ & FN & R \\
\midrule
Import & 785 & 233 & 0.703 \\
File/path & 311 & 82 & 0.736 \\
Symbol/export & 243 & 48 & 0.802 \\
Tool/harness schema & 49 & 15 & 0.694 \\
Environment & 13 & 12 & 0.077 \\
\midrule
Total & 1,401 & 390 & 0.722 \\
\bottomrule
\end{tabular}
\end{minipage}\hfill
\begin{minipage}[t]{0.3\textwidth}
\centering
\scriptsize
\textbf{False-negative layer}\\[2pt]
\begin{tabular}{@{}lr@{}}
\toprule
Layer & Count \\
\midrule
Code extraction & 106 \\
Command extraction & 95 \\
Selected-file reconstruction & 87 \\
Weakly typed target & 51 \\
Dependency closure & 21 \\
Indirect/reconstruction & 14 \\
State probe/guard & 13 \\
Pytest collection closure & 3 \\
\midrule
Total & 390 \\
\bottomrule
\end{tabular}
\end{minipage}
\end{table*}

%% file: Sections/RQs/RQ2.tex
\subsection{RQ5: Cost Analysis}

A pre-execution detector must maintain strict efficiency because it evaluates every submitted action, whereas only a fraction actually contain hallucinated tool calls. If this per-action overhead approaches the cost of a runtime error, the intervention provides no net benefit. We therefore evaluate the latency and token cost each method imposes per submitted action.
% For \tool, the timer encompasses predicate extraction and deterministic state probing, including repository reads and AST inspections.
% For the Action-only baseline, it measures extraction alone, whereas the scaffold linter measures AST snippet linting.

\input{Tables/tab-rq2-latency}
Table~\ref{tab:rq2-latency} compares the runtime latency and token overhead across approaches. 
% but differ fundamentally in precision.
% The scaffold linter executes in just 0.004\,ms at the median but checks only local syntax without repository grounding, yielding a negligible 0.045 F1.
% Action-only extraction requires 0.75\,ms at the median, yet its ungrounded checks generate 995 false positives.
\tool incurs a median latency of only 1.50\,ms (306.66\,ms at p95) with zero token cost for detection. 
This demonstrates that workspace probing provides a fast and necessary grounding for effective pre-execution detection.
LLM judges, however, incur substantial inference latency and token usage on every audited action.
Their median latencies range from 14.51\,s to 17.73\,s, which is over four orders of magnitude slower than \tool.
% Maximum response times for LLMs can even reach several minutes due to network and provider delays.
Furthermore, evaluating the benchmark with LLM judges consumes between 1.9K and 19.2K tokens per action, regardless of whether the action contains a hallucinated tool call.
This continuous overhead makes LLM-based auditing impractical in interactive trajectories. 

% In contrast, \tool consumes no tokens during detection; it only incurs a token cost when an agent must read a warning and regenerate a replacement for a detected hallucination at runtime.

\begin{tcolorbox}[
    colback = sub,
    colframe = main,
    boxrule = 0pt,
    toprule = 1pt, % top rule weight
    bottomrule = 1pt, % bottom rule weight
    sharp corners
]
\vspace{-0.5em}
\small
\label{box:rq2}
\textbf{Answer to RQ5:} 
\tool operates token-free with a median latency of 1.50\,ms, reducing detection latency from seconds to milliseconds compared with LLM judges.
\vspace{-0.5em}
\end{tcolorbox}

%% file: Tables/tab-rq2-latency.tex
\begin{table}[t]
\centering
\caption{RQ5: Statistics of pre-execution latency and token consumption.
Tokens denote aggregate input and output tokens across all evaluated actions.}
\label{tab:rq2-latency}
\small
\begin{tabular}{lrrrrrr}
\toprule
\textbf{Method} & \textbf{Mean} & \textbf{Median} & \textbf{p95} & \textbf{p99} & \textbf{Max} & \textbf{Tokens} \\
\midrule
Scaffold linter & 0.043 ms & 0.004 ms & 0.220 ms & 0.963 ms & 10.28 ms & 0 \\
Action-only & 25.25 ms & 0.75 ms & 16.57 ms & 310.03 ms & 2.67 s & 0 \\
\tool (Ours)& 63.22 ms & 1.50 ms & 306.66 ms & 1.09 s & 11.73 s & 0 \\
\midrule
Qwen3.7 Plus & 25.89 s & 17.73 s & 54.01 s & 199.81 s & 325.90 s & 4.87M \\
Kimi K3 & 25.20 s & 16.80 s & 62.85 s & 188.98 s & 668.56 s & 4.84M \\
Codex 5.6 Sol & 15.87 s & 14.51 s & 28.45 s & 34.05 s & 54.23 s & 48.15M \\
\bottomrule
\end{tabular}
\end{table}

%% file: Sections/RQs/RQ3.tex
\subsection{RQ6: Ablation Analysis}
To evaluate the individual contributions of different predicate families, we perform both cumulative addition and leave-one-out ablation studies of \tool.
Starting from core namespace and symbol checks, we incrementally introduce and systematically isolate each predicate family to quantify its marginal impact on precision, recall, F1-score and accuracy.

\input{Tables/tab-rq3-ablation}
Table~\ref{tab:rq3-ablation} summarizes the ablation results.
Namespace and package-layout predicates (Family~I) provide the high-precision foundation, identifying 752 true positives with 0.972 precision and 0.691 F1; correspondingly, omitting this family in the leave-one-out setting causes the sharpest degradation, collapsing F1 to 0.323.
Filesystem and cwd predicates (Family~II) produce the primary recall expansion, recovering 224 additional true positives and raising recall from 0.537 to 0.697 (0.801 F1), while removing them reduces F1 to 0.711.
Harness and tool-schema predicates (Family~IV) contribute 34 additional true positives at the cost of three false positives, elevating overall F1 to 0.817 (and removing them lowers F1 to 0.801).
Command, test, and runtime predicates (Family~III) add a single true positive, an absent \texttt{pytest} node, because this benchmark contains few command-level hallucinations.
Overall, filesystem and cwd predicates are the primary coverage extension beyond the high-precision namespace core, while the remaining families offer complementary domain-specific gains.

\begin{tcolorbox}[
    colback = sub,
    colframe = main,
    boxrule = 0pt,
    toprule = 1pt, % top rule weight
    bottomrule = 1pt, % bottom rule weight
    sharp corners
]
\vspace{-0.5em}
\small
\label{box:rq3}
\textbf{Answer to RQ6:} \tool relies on namespace and package-layout predicates as its high-precision core, while filesystem and cwd predicates raise F1 from 0.691 to 0.801, with remaining families providing complementary coverage up to 0.817 F1.
\vspace{-0.5em}
\end{tcolorbox}

%% file: Tables/tab-rq3-ablation.tex
\begin{table}[t]
\centering
\caption{RQ6: Statistics of cumulative and leave-one-out ablations over the four predicate families (I--IV).}
%Accuracy is $(\mathrm{TP}+\mathrm{TN})/2508$.
\label{tab:rq3-ablation}
\small
\setlength{\tabcolsep}{2.5pt}
\begin{tabular}{@{}lrrrrrrrr@{}}
\toprule
\textbf{Variant} & \textbf{TP} & \textbf{FP} & \textbf{FN} & \textbf{TN} &
\textbf{P} & \textbf{R} & \textbf{F1} & \textbf{Acc} \\
\midrule
Full & 1011 & 64 & 390 & 1043 & 0.940 & 0.722 & 0.817 & 0.819 \\
\midrule
\multicolumn{9}{@{}l}{\textbf{Cumulative addition}} \\
I: Namespace/package & 752 & 22 & 649 & 1085 & 0.972 & 0.537 & 0.691 & 0.732 \\
+ II: Filesystem/cwd & 976 & 61 & 425 & 1046 & 0.941 & 0.697 & 0.801 & 0.806 \\
+ III: Command/test/runtime & 977 & 61 & 424 & 1046 & 0.941 & 0.697 & 0.801 & 0.807 \\
+ IV: Harness/tool schema & 1011 & 64 & 390 & 1043 & 0.940 & 0.722 & 0.817 & 0.819 \\
% \addlinespace[2pt]
\midrule
\multicolumn{9}{@{}l}{\textbf{Leave one predicate family out}} \\
w/o I: Namespace/package & 278 & 42 & 1123 & 1065 & 0.869 & 0.198 & 0.323 & 0.535 \\
w/o II: Filesystem/cwd & 787 & 25 & 614 & 1082 & 0.969 & 0.562 & 0.711 & 0.745 \\
w/o III: Command/test/runtime & 1010 & 64 & 391 & 1043 & 0.940 & 0.721 & 0.816 & 0.819 \\
w/o IV: Harness/tool schema & 977 & 61 & 424 & 1046 & 0.941 & 0.697 & 0.801 & 0.807 \\
\bottomrule
\end{tabular}
\end{table}

%% file: Sections/RQs/RQ4.tex
\input{Tables/tab-rq4-mh-sensitivity}
\subsection{RQ7: Sensitivity Analysis}
\label{sec:rq4}
To evaluate whether tracing repository-local dependencies beyond direct predicates improves detection, we investigate bounded multi-hop probing across the 2,508 benchmark actions.
Specifically, beyond the direct imports checked by default, \tool-\texttt{MH1} and \tool-\texttt{MH2} additionally check transitive repository-local imports up to one and two further levels, respectively.
% while non-local boundaries such as standard libraries, dynamic imports, and reflection conservatively propagate \texttt{unknown}.
% All evaluation labels and intervention thresholds remain fixed, isolating the direct effect of transitive probing depth.

As reported in Table~\ref{tab:rq2-mh-sensitivity}, deeper static probing yields negligible coverage improvements while introducing a disproportionate precision and latency penalty.
Compared to the default configuration, \texttt{MH1} recovers five additional true positives at the cost of five false positives, leaving F1 at 0.817 and accuracy at 0.819, whereas \texttt{MH2} recovers eight true positives but incurs eleven false positives (F1 stays at 0.817 and accuracy falls to 0.818).
Furthermore, \texttt{MH1} and \texttt{MH2} take $1.65\times$ and $2.68\times$ the p95 tail latency of the default version.
These additional false positives predominantly stem from deep package surfaces where conditional imports, compatibility aliases, and re-export hubs weaken the localization of static checks.
Consequently, checking only direct imports provides the optimal trade-off for pre-execution intervention, while multi-hop evidence is best reserved for non-blocking diagnostic modes.
\begin{tcolorbox}[
    colback = sub,
    colframe = main,
    boxrule = 0pt,
    toprule = 1pt, % top rule weight
    bottomrule = 1pt, % bottom rule weight
    sharp corners
]
\vspace{-0.5em}
\small
\label{box:fd4}
\textbf{Answer to RQ7:} Extending \tool to multi-hop static probing provides negligible coverage improvements while introducing a higher rate of false positives and increasing p95 tail latency.
\vspace{-0.5em}
\end{tcolorbox}
% Direct, one-hop checks offer the optimal trade-off for conservative pre-execution gating.

%% file: Tables/tab-rq4-mh-sensitivity.tex
\begin{table}[t]
\centering
\caption{RQ7: Sensitivity analysis across multi-hop probing depths. Latency is reported in milliseconds.
% Accuracy is $(\mathrm{TP}+\mathrm{TN})/2508$.
Bold and underline indicate the best and second-best results, respectively.}
\label{tab:rq2-mh-sensitivity}
% \small
\setlength{\tabcolsep}{2.2pt}
\begin{tabular}{lrrrrrrrrrr}
\toprule
\textbf{Variant} & \textbf{TP} & \textbf{FP} & \textbf{FN} & \textbf{TN} &
\textbf{P} & \textbf{R} & \textbf{F1} & \textbf{Acc} & \textbf{Med.} & \textbf{p95} \\
\midrule
\tool (default) & 1011 & \textbf{64} & 390 & \textbf{1043} & \textbf{0.940} & 0.722 & \textbf{0.817} & \textbf{0.819} & \textbf{1.499} & \textbf{306.662} \\
\tool-MH1 & \underline{1016} & \underline{69} & \underline{385} & \underline{1038} & \underline{0.936} & \underline{0.725} & \textbf{0.817} & \textbf{0.819} & \underline{2.342} & \underline{506.356} \\
\tool-MH2 & \textbf{1019} & 75 & \textbf{382} & 1032 & 0.931 & \textbf{0.727} & \textbf{0.817} & \underline{0.818} & 2.867 & 822.877 \\
\bottomrule
\end{tabular}
\end{table}

%% file: Sections/RQs/RQ5.tex
\subsection{RQ8: Mitigation of Hallucinated Tool Calls}
\label{sec:rq5}

\begin{figure*}[t]
    \centering
    \includegraphics[width=\textwidth]{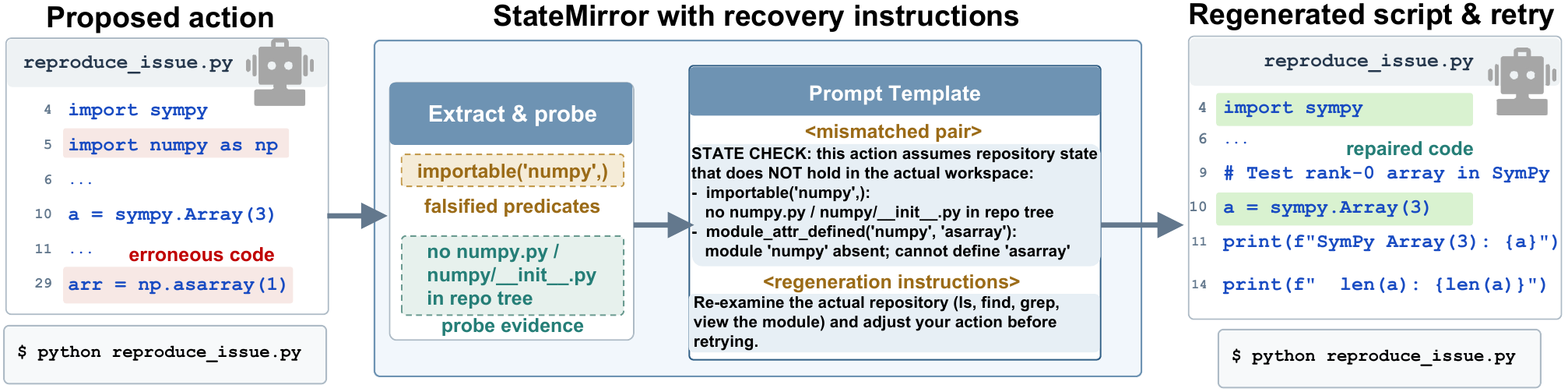}
    \caption{RQ8: An example of live action repair and pre-execution intervention.}
    \label{fig:rq5-live-repair}
\end{figure*}

We investigate how effectively \tool mitigates detected hallucinated tool calls within agent workflows, at two levels with recovery instructions (as illustrated in Figure~\ref{fig:rq5-live-repair}).
\emph{Action repair} asks whether a typed warning lets the agent replace the faulty action with a valid one, under both controlled revision and live agent execution.
\emph{Pre-execution intervention} asks whether blocking the faulty action before it runs removes the downstream cost the hallucination would otherwise cause over the rest of the task.
We run \tool in two modes: \emph{report mode}, which logs every warning without blocking, and \emph{intervention mode}, which applies a conservative blocking rule.
In intervention mode, \tool blocks an action and prompts regeneration only if a falsified predicate is a necessary precondition, such as an executable or a file to be read.
It permits predicates related to intentional side effects (e.g., files the action aims to create) and diagnostic errors (e.g., missing dependencies within existing test files), blocking test commands only when the test target or runner itself is missing.

\input{Tables/tab-rq5-actionability}
\subsubsection{Action repair.}
To evaluate whether agents can correct pending errors given pre-execution feedback, we perform controlled action revision on 100 hallucinated tool calls (sampled evenly across the four agent sources).
Each agent receives its pre-action workspace context, the failed action, and \tool's warning, without any execution feedback.
As shown in Table~\ref{tab:rq5-actionability}, agents achieve an overall repair rate of 95.0\%. 
Repair rates reach 100\% on Honeycomb and SWE-agent, 96.0\% on OpenHands, and 84.0\% on Sonar.
Of the five unrepaired cases, one repeats the warned predicate, and four replace the missing symbol with a sibling name that the package does not export either.
Crucially, 88 of the 95 replacements introduce no secondary hallucinations, while only seven surface a new mismatch.
Pre-execution detection of these replacements also adds minimal overhead (median 21.5\,ms).

To evaluate repair in an active agent loop, we deploy \tool with SWE-agent across 385 SWE-bench Verified tasks.
In report mode, \tool audits 3,829 actions and emits 50 warnings across 49 tasks, 46 of which are confirmed by immediate execution failures.
Rerunning these 49 tasks in intervention mode yields 46 evaluable events where the hallucination reproduces and triggers blocking.
The agent's very first replacement clears the mismatch and executes successfully in 45 of these 46 cases, achieving a 97.8\% live repair rate.
This confirms that deterministic pre-execution warnings are highly actionable, allowing agents to preempt execution failures and recover in a single turn.
% \input{Tables/tab-rq5-live-repair}

% The run records two unnecessary intervention events, four later same-subject
% rehits, zero correction-budget exhaustions, zero StateMirror-induced
% terminations, and zero hook errors.

\input{Tables/tab-rq5-task-resolution}
\subsubsection{Pre-execution intervention.}
We next evaluate whether pre-execution intervention mitigates the downstream cost of hallucinated tool calls over the remainder of the task.
From 475 Qwen3.8 SWE-agent trajectories run in report mode, \tool detects actionable hallucinations in 79 runs.
For these 79 cases, we replay the original trajectory up to the detected action and then let the agent continue in intervention mode, ensuring both arms share the exact same prefix.
All runs adhere to a strict 100-step ceiling, with each intervention warning and replacement generation counting against the budget to ensure a fair comparison.
Table~\ref{tab:rq5-task-resolution} reports how the action is repaired and how much downstream cost the intervention removes.

In the intervention cohort, the agent produces a replacement in 78 of the 79 cases before exhausting the budget. The first replacement clears every detected predicate in 73 cases (93.6\%), and 66 of those (84.6\%) successfully execute with exit code 0. Most non-clears simply repeat the missing dependency through a different command (e.g., an uninstalled \texttt{numpy}). Compared with report mode (739.2\,s, 56.0 steps, 805.0K tokens per case), the full intervention trajectories average 622.7\,s, 52.7 steps, and 709.9K tokens. This yields reductions of 15.8\%, 6.0\%, and 11.8\%, saving roughly 116.5\,s, 3.4 steps, and 95.1K tokens per triggered task. Importantly, these savings do not come at the expense of task success: both modes resolve exactly 48 of the 79 tasks.

These savings represent the direct cost of the hallucinated tool calls. They correspond to the wasted execution time, error observations, and recovery turns that pre-execution intervention successfully avoids. Because \tool itself adds no tokens, every saved step becomes a step available for genuine repair under the fixed 100-step budget ceiling. While absolute agent-time savings depend on provider latency and queue scheduling, the reductions in tokens and steps confirm that deterministic intervention improves execution efficiency.

\begin{tcolorbox}[
    colback = sub,
    colframe = main,
    boxrule = 0pt,
    toprule = 1pt, % top rule weight
    bottomrule = 1pt, % bottom rule weight
    sharp corners
]
\vspace{-0.5em}
\small
\label{box:fd5}
\textbf{Answer to RQ8:} \tool effectively mitigates hallucinated tool calls. It enables successful action repair in 95.0\% of controlled cases and 97.8\% of live executions. This live intervention reduces agent time by 15.8\%, execution steps by 6.0\%, and token consumption by 11.8\% per triggered task, without degrading end-to-end task resolution.
\vspace{-0.5em}
\end{tcolorbox}

%% file: Tables/tab-rq5-actionability.tex
\begin{table}[t]
\centering
\caption{RQ8: Controlled action revision performance across agent sources and hallucination categories.}
\label{tab:rq5-actionability}
\scriptsize
\setlength{\tabcolsep}{4pt}
\begin{tabular*}{\columnwidth}{@{\extracolsep{\fill}}lrrrrrrrr@{}}
\toprule
& \multicolumn{5}{c}{\textbf{Source}} & \multicolumn{3}{c}{\textbf{Hallucination category}} \\
\cmidrule(lr){2-6}\cmidrule(lr){7-9}
& \textbf{All} & \textbf{Honeycomb} & \textbf{OpenHands} & \textbf{Sonar} & \textbf{SWE-agent}
& \textbf{File} & \textbf{Import} & \textbf{Symbol} \\
\midrule
Repaired & 95 & 25 & 24 & 21 & 25 & 27 & 44 & 24 \\
$N$ & 100 & 25 & 25 & 25 & 25 & 27 & 48 & 25 \\
Rate & 0.950 & 1.000 & 0.960 & 0.840 & 1.000 & 1.000 & 0.917 & 0.960 \\
\bottomrule
\end{tabular*}
\end{table}

%% file: Tables/tab-rq5-task-resolution.tex
% Evidence (action repair): artifacts/analysis/rq5_qwen38_replay_repeats2_20260910/action_repair_summary.json (round01; action_repair_summary.py)
% Evidence (resolution): artifacts/analysis/rq5_qwen38_replay_repeats2_20260910/final_three_round_summary.json (rounds.1)
% Evidence (efficiency): artifacts/analysis/rq5_qwen38_replay_repeats2_20260910/full_logical_efficiency_summary.json
\begin{table}[t]
\centering
\caption{RQ8: Task-level hallucination mitigation on 79 StateMirror-triggered cases from Qwen3.8 SWE-agent runs.
\emph{Action repair} evaluates the first replacement after a warning; \emph{cleared} means no warned predicate recurs, and \emph{executed} requires exit code 0.
Time, steps, and tokens represent the average consumption per case. }
\label{tab:rq5-task-resolution}
% \small
\setlength{\tabcolsep}{5pt}
\begin{tabular}{@{}lrrrrrr@{}}
\toprule
& \multicolumn{2}{c}{\textbf{Action repair} ($N=78$)}
& \textbf{Agent time} & \textbf{Steps} & \textbf{Tokens} & \textbf{Resolved} \\
\cmidrule(lr){2-3}
\textbf{Mode} & Cleared & + Executed & (s) & & ($\times 10^3$) & (/79) \\
\midrule
Report & -- & -- & 739.2 & 56.0 & 805.0 & 48 \\
Intervention & 73 (93.6\%) & 66 (84.6\%) & 622.7 & 52.7 & 709.9 & 48 \\
\midrule
$\Delta$ & & & $\mathbf{-15.8\%}$ & $\mathbf{-6.0\%}$ & $\mathbf{-11.8\%}$ & $\pm 0$ \\
\bottomrule
\end{tabular}
\end{table}

%% file: Sections/06-case-study.tex
\begin{figure}[t]
    \centering
    \includegraphics[width=\textwidth]{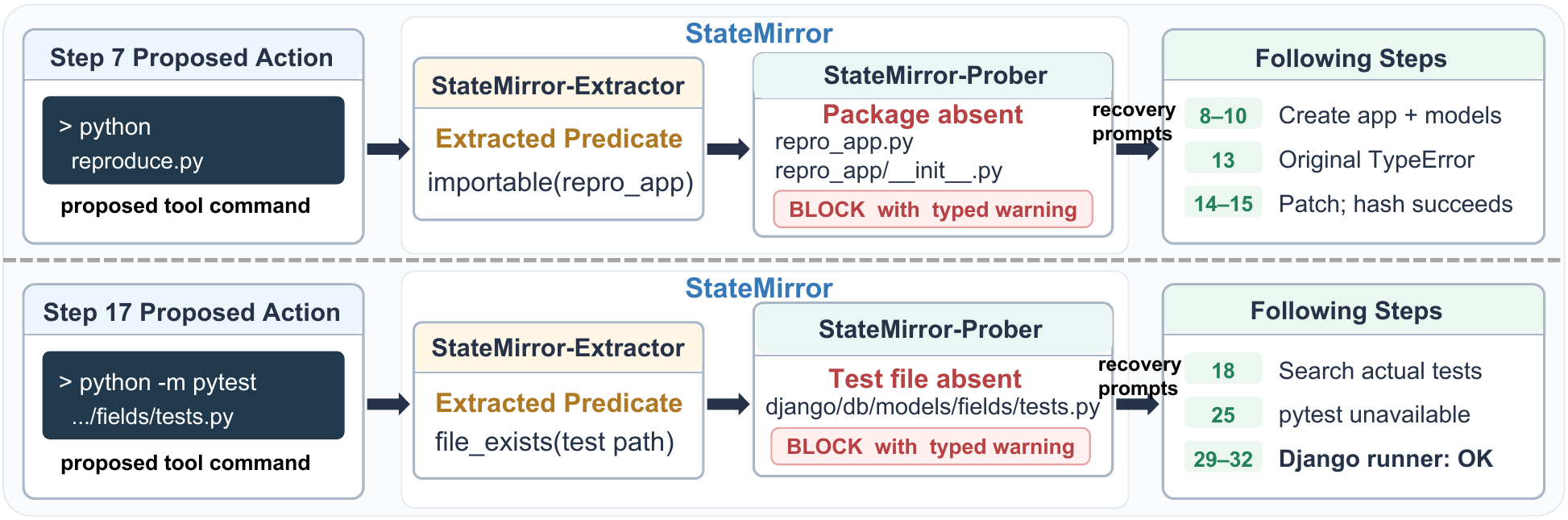}
    \caption{A runtime detection example in \texttt{django\_\_django-14672}.
    At steps 7 and 17, \tool blocks commands that assume a missing reproduction
    package and test file. The agent continues to a resolved patch.}
    \Description{Two horizontal action-to-effect workflows from one Django
    coding-agent run. In the first, step 7 submits a reproduction command;
    StateMirror extracts a required import, probes the missing repro app, and
    blocks execution. The agent creates the package, reproduces the hash error,
    then patches the source. In the second, step 17 targets an absent test
    file; StateMirror blocks that command. The agent searches for tests and
    later runs Django's own test runner successfully.}
    \label{fig:case-runtime-django}
\end{figure}

\section{Case Study}
We conduct a case study to show how \tool detects hallucinated tool calls in practice. Specifically, we trace a live SWE-agent run on \texttt{django\_\_django-14672} from the mitigation study. The task concerns hashing a \texttt{ManyToManyRel} whose \texttt{through\_fields} is a list. Figure~\ref{fig:case-runtime-django} shows two pre-execution interventions and the subsequent actions observed in the same trajectory.

At step 7, the agent submits \texttt{python reproduce.py}, which imports a non-existent \texttt{repro\_app}; \tool flags \texttt{importable(repro\_app)} as false and halts execution, guiding the agent to create the missing definitions before reproducing the issue and patching \texttt{ManyToManyRel.identity}.
Later at step 17, when the agent attempts to run tests via \path{django/db/models/fields/tests.py}, \tool intercepts the non-existent test path with a \texttt{file\_exists} warning, prompting the agent to inspect the test suite and switch to the valid test runner \texttt{tests/runtests.py}, ultimately leading to a resolved patch.

This case study highlights that \tool serves as an effective fail-fast safeguard in live trajectories.
By intercepting invalid environment assumptions before execution, \tool prevents uninformative runtime failures, reduces wasted recovery cost, and guides the agent toward valid workspace state, effectively mitigating hallucinated tool calls and enabling continuous progress.
% An independent
% report-mode run also resolves this task, so this trace establishes local
% pre-execution corrections and subsequent success, not that the warnings caused
% the task-level outcome.

%% file: Sections/related-work.tex
\section{Related Work}
% \subsection{Coding Agent Auditing and Debugging}
\textbf{Coding Agent Auditing and Debugging.} As interactive coding agents grow more popular, researchers have developed various auditing, debugging, and post-execution repair techniques~\cite{wang2026trajaudit, wang2026agent}.
Traditional debugging and automated repair methods analyze failures after program execution to locate faults or suggest patches~\cite{zeller2002delta,liblit2005statistical,jones2005tarantula,wong2016survey,fan2023llmrepair,jiang2023impact}.
Similarly, feedback-driven agent systems run commands and use the resulting runtime tracebacks to correct their actions~\cite{shinn2023reflexion,chen2023selfdebug}.
Other trajectory analysis frameworks evaluate multi-turn loops to identify failure patterns~\cite{liu2023agentbench,ma2024agentboard,bouzenia2025trajectories,cemri2025multiagentfail}, or analyze execution trajectories and their prefixes for failure attribution, diagnosis, online detection, and repair~\cite{zhang2025whowhen,zhu2025agentdebug,barke2026agentrx,zhang2026agentforesight,chen2026harnessfix}.
% However, these approaches require the agent to run the erroneous action first, which often wastes execution steps, consumes excessive tokens, and increases monetary costs.
While AgentSpec~\cite{wang2025agentspec} can perform pre-execution checks using rules derived from user-specified safety requirements, StateMirror derives factual preconditions from the proposed tool call itself and verifies them through deterministic probes of the current execution environment.

The post-hoc limitation is particularly evident when dealing with hallucinations in generated code.
In language models, hallucination represents output that is unfaithful to a source or to the real world~\cite{huang2025hallucinationsurvey}.
For code generation tasks, these errors often remain hidden until the program actually runs.
For example, models frequently generate code that references unpublished packages~\cite{spracklen2025packagehallucination} or fails during execution~\cite{zhang2025codehallucination,wang2025codegenerrors,tian2025codehalu}.
While existing methods analyze or mitigate errors in finished code snippets~\cite{jain2025apidoc}, they do not check whether the required modules, symbols, and paths are present in the target workspace before running the agent's next tool call.
% To address these challenges, \tool operates at the pre-execution boundary.
Post-execution systems generally run actions first and resolve errors using tracebacks, which often increases execution steps and token consumption.
In contrast, \tool serves as a proactive gate that verifies repository requirements with lightweight, read-only state probes before execution.
% This approach prevents unnecessary execution steps and reduces token overhead, making it a valuable complement to existing post-execution tools.

%% file: Sections/07-threats.tex
\section{Threats to Validity}

\textbf{Internal threats.}
To ensure internal validity, we mitigate annotation subjectivity by establishing benchmark labels through independent expert annotations and consensus adjudication.
We reconstruct workspace states with high fidelity by replaying verified prefix edits.
Furthermore, our evaluation includes both failed and resolved trajectories across four different agent frameworks to minimize framework-specific bias.
On the resolved stratum, we observe a lower recall because some errors involve dynamic runtime states rather than static mismatches.
Currently, our deterministic predicates prioritize actionable, high-precision static checks.
Extending this validation to dynamic, execution-time errors represents an important avenue for future work.

\textbf{External threats.} While our current empirical evaluation is instantiated on Python bug-fixing tasks, SWE-agent trajectories, and the Qwen3.8-27B model, the \tool architecture is designed for broad generalizability. Because it verifies foundational execution assumptions (e.g., filesystem and namespace states) rather than relying on bug-fixing-specific heuristics, its core formulation naturally extends to other coding agents such as code-generation agents and testing agents.
Generalizing \tool to new programming languages or agent runtimes involves implementing domain-specific predicate extractors and probers, while the overall pre-execution detection workflow remains intact.

%% file: Sections/08-Conclusion.tex
\section{Conclusion}
Hallucinated tool use occurs when an agent's action assumes invalid workspace preconditions. Our preliminary study of 4,398 trajectories reveals these errors are highly prevalent, expensive to recover, and concentrated in filesystem and namespace mismatches.
To address this, we present \tool, a lightweight pre-execution detector that extracts predicates from actions and deterministically validates them against the live environment.
Evaluation on 2,508 actions shows \tool achieves 0.940 precision and 0.817 F1, outperforming the strongest LLM judge by 35.3\% in F1 with zero token cost and millisecond latency.
For practical mitigation, \tool enables agents to rectify over 95\% of hallucinated tool calls, reducing execution time by 15.8\% and token consumption by 11.8\% during live intervention.
% Hallucinated tool use arises when a submitted action assumes workspace preconditions that do not hold. Agents typically recover only after a runtime error, which wastes execution steps and tokens.
% Our preliminary study of 4,398 SWE-agent trajectories shows that such hallucinations appear in over 30\% of both successful and failed tasks, and that each occurrence costs a median of 3 agent steps and 22.4K tokens.
% We present \tool, a lightweight pre-execution detector that intercepts actions at the harness boundary, extracts typed state predicates, and validates them against the live repository with read-only probes.
% A comprehensive evaluation on 2,508 labeled actions shows that \tool achieves high detection accuracy, with 0.940 precision and 0.817 F1 at a 1.50-ms median latency and zero tokens, exceeding the strongest LLM judge by 35.3\% in F1.
% To show the practicality of \tool{} in mitigating hallucinated tool calls, we evaluate both action repair and pre-execution intervention. \tool enables action repair in 95.0\% of controlled cases and 97.8\% of live executions, and its pre-execution intervention reduces agent time by 15.8\% and token consumption by 11.8\%.